\documentclass{emulateapj}

\shorttitle{Kinematic Decoupling of EHB GCs}
\shortauthors{Lee, Gim, \& Casetti-Dinescu}

\submitted{Accepted for Publication in ApJ Letters, April 2007}

\begin{document}
\title { KINEMATIC DECOUPLING OF GLOBULAR CLUSTERS \\
               WITH EXTENDED HORIZONTAL-BRANCH}
\author{Young-Wook Lee\altaffilmark{1,2}, Hansung B. Gim\altaffilmark{1}, \& Dana I. Casetti-Dinescu\altaffilmark{2}}
\altaffiltext{1} {Center for Space Astrophysics $\&$ Department of Astronomy, Yonsei University, Seoul 120-749, Korea (ywlee@csa.yonsei.ac.kr)}
\altaffiltext{2} {Department of Astronomy, Yale University, New Haven, CT 06520, USA}

\begin{abstract}

About 25$\%$ of the Milky Way globular clusters (GCs) exhibit unusually extended color distribution of stars in the core helium-burning 
horizontal-branch (HB) phase. This phenomenon is now best understood as due to the presence of helium enhanced second generation 
subpopulations, which has raised a possibility that these peculiar GCs might have a unique origin. Here we show that these 
GCs with extended HB are clearly distinct from other normal GCs in kinematics and mass. The GCs with extended HB are 
more massive than normal GCs, and are dominated by random motion with no correlation between kinematics and metallicity. 
Surprisingly, however, when they are excluded, most normal GCs in the inner halo show clear signs of dissipational collapse 
that apparently led to the formation of the disk. Normal GCs in the outer halo share their kinematic properties with 
the extended HB GCs, which is consistent with the accretion origin. Our result further suggests heterogeneous origins of GCs, 
and we anticipate this to be a starting point for more detailed investigations of Milky Way formation, 
including early mergers, collapse, and later accretion.
\end{abstract}


\section{INTRODUCTION}
The discovery of multiple stellar populations in the most massive GC \object{$\omega$Cen} (\citealp{Lee99}), together with the fact 
that the second most massive GC \object{M54} is a core of the disrupting Sagittarius dwarf galaxy (\citealp{LS00}), 
have strengthen the view that some of the massive GCs might be remaining cores of disrupted nucleated dwarf galaxies (\citealp{Freeman93}). 
Among their several peculiar characteristics, both $\omega$Cen and M54 have extended horizontal-branch (EHB), 
with extremely hot horizontal-branch (HB) stars well separated from redder HB (\citealp{Lee99}; Rosenberg, Recio-Blanco, $\&$ Garcia-Marin 2004). 
High precision $Hubble$ $Space$ $Telescope$ ($HST$) photometry (\citealp{Bedin04}) has discovered that $\omega$Cen also has a curious 
double main-sequence (MS). Recent studies have shown that both these peculiar colour-magnitude diagram (CMD) 
characteristics are best understood as due to the presence of helium enhanced second generation 
subpopulations (\citealp{Norris04}; \citealp{Lee05}; \citealp{Piotto05}; \citealp{Dan05}). Furthermore, 
the prediction of the models (\citealp{Lee05}; \citealp{Dan05}) that most of the GCs with EHB would have double 
or broadened MSs are now confirmed by $HST/ACS$ ($Advanced$ $Camera$ $for$ $Survey$) photometry (\citealp{Piotto07}). 
This ensures that EHBs are strong signature of the presence of multiple populations in GCs. A significant fraction ($\sim$30$\%$) 
of the helium enriched subpopulation observed in these peculiar GCs is also best explained if the second generation stars 
were formed from enriched gas trapped in the deep gravitational potential well while these GCs were cores of 
the ancient dwarf galaxies (\citealp{BN06}). Despite the lack of apparently wide spread in iron-peak elements 
in most of these GCs, all of these recent developments suggest that GCs with EHB are probably not genuine GCs, 
but might have a unique origin in the formation history of the Galaxy.

In order to test this working hypothesis further, we have carefully surveyed 114 GCs with reasonably good CMDs, 
and found that 28 (25$\%$) of them have EHB (Lee et al., in preparation). Their NGC numbers are: \object{2419}, \object{2808}, \object{5139}, \object{5986}, \object{6093}, \object{6205}, \object{6266}, \object{6273}, 
\object{6388}, \object{6441}, \object{6656}, \object{6715}, \object{6752}, \object{7078}, and \object{7089} for the GCs with strongly extended HB; and \object{1851}, \object{1904}, \object{4833}, \object{5824}, \object{5904}, 
\object{6229}, \object{6402}, \object{6522}, \object{6626}, \object{6681}, \object{6712}, \object{6723}, and \object{6864} for the GCs with moderately extended HB, 
including those with bimodal HB distributions. We will collectively call all of them as ``EHB GCs''. 
Our selection of EHB GCs were based on the reddening independent criteria on CMD in \textit{B}$\&$\textit{V} passbands ($\Delta$$V_{HB}$ $>$ 3.5 
for strongly extended HB; either 3.0 $<$ $\Delta$$V_{HB}$ $<$ 3.5 or $\Delta$$(B-V)_{HB}$ $>$ 0.78 with clear bimodal colour 
distribution for moderately extended HB). But, since their appearances on CMDs are distinct enough from GCs 
with normal HB (\citealp{Piotto02}), our selection agrees well with the result based on smaller sample and other measures of 
HB temperature extension (e.g., \citealp{RB06}). We have then investigated their properties compared to other normal GCs.

\section{LUMINOSITY FUNCTION AND KINEMATICS}

First of all, from the luminosity function (Fig. 1), we found that EHB GCs are among the brightest GCs of the Milky Way, 
including 11 out of 12 brightest GCs (see also \citealp{RB06}). It is surprising to see that not a single EHB GC is fainter than $M_{V}$ = -7. 
Careful inspection of all CMDs confirms that this is not due to the smaller number of HB stars in fainter GCs. 
Because of significant fraction (18 - 51$\%$) of the helium enriched bluer subpopulation observed in EHB GCs, 
its presence on the HB would be reliably detected ($>$ 5 - 10 stars) even in a cluster of $M_{V}$ = -6 or -5 if it existed. 
This result is perhaps already suggesting that EHB GCs might have a peculiar origin, as their inferred current stellar mass, 
which might represent only a small fraction of their original mass, is comparable with that of low-luminosity 
dwarf galaxies in the Local Group.

\begin{figure}
\epsscale{0.90}
\plotone{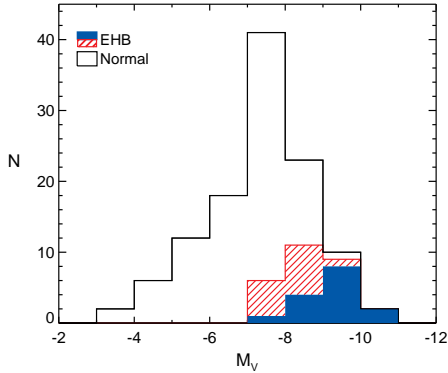}
\caption{The histogram of $M_{V}$ for 114 Milky Way GCs (data from \citealp{Harris96}). 
Blue and red are GCs with strongly and moderately extended HBs, respectively. EHB GCs are clearly brighter (more massive) than normal GCs.}
\end{figure}

Motivated by this, we have investigated the kinematics of EHB GCs, in order to see whether their kinematic properties 
are also distinct from other normal GCs. Following previous investigation (\citealp{Zinn93}), we have first divided GCs into 
three subgroups (Fig. 2) based on the HB morphology and metallicity diagram (Lee, Demarque, $\&$ Zinn 1994). 
Metal-poor ([Fe/H] $<$ -0.8) GCs in the ``Old halo (OH)'' group have bluer HB morphology at a given metallicity, 
and those in the ``Younger halo (YH)'' group have redder HB morphology at fixed metallicity. 
The OH group, in the mean, is probably older than the YH group by $\sim$1 Gyr (\citealp{Rey01}; \citealp{SW02}). 
The metal-rich ([Fe/H] $>$ -0.8) GCs are further classified as ``disk/bulge (D/B)'' group. EHB GCs belong in all 
three subgroups, although the majority of them are in the OH group. Note also that most (94$\%$) 
GCs in YH group are in the outer halo (galactocentric distance, $R_{gc}$ $>$ 8 kpc), 
while the majority (80$\%$) of GCs in OH and D/B groups are in the inner halo ($R_{gc}$ $<$ 8 kpc).

\begin{table}
\begin{center}
\caption{KINEMATICS OF GLOBULAR CLUSTERS BASED ON RADIAL VELOCITY ALONE\tablenotemark{*}}
\label{tab:columns}
\setlength{\tabcolsep} {9pt}
  \begin{tabular}{@{}cccccc@{}} 
   \hline\hline
   Group & N & $V_{rot}$ & $\sigma_{los}$ & $V_{rot}$/$\sigma_{los}$\\
\tableline
   \multicolumn{2}{l}{All GCs}      &            &            &             \\ 
     All Halo & 71 & 25$\pm$27  & 124$\pm$10 & 0.20$\pm$0.22\\
     YH       & 25 & -18$\pm$66 & 153$\pm$22 & -0.12$\pm$0.43\\
     OH       & 46 & 40$\pm$27  & 104$\pm$11 & 0.38$\pm$0.26\\
     D/B      & 14 & 168$\pm$28 & 65$\pm$12  & 2.57$\pm$0.65\\
\tableline
   \multicolumn{2}{l}{EHB GCs}      &            &            &             \\
     All EHB  & 24 & 10$\pm$32  & 93$\pm$13  & 0.11$\pm$0.34\\
     OH       & 18 & 4$\pm$35   & 91$\pm$15  & 0.05$\pm$0.38\\
\tableline
   \multicolumn{2}{l}{Normal}       &            &            &              \\
     All Halo & 48 & 32$\pm$39  & 137$\pm$14 & 0.24$\pm$0.29\\
     YH       & 20 & -42$\pm$80 & 162$\pm$26 & -0.26$\pm$0.49\\
     OH       & 28 & 70$\pm$39  & 111$\pm$15 & 0.63$\pm$0.36\\
     D/B      & 13 & 188$\pm$22 & 48$\pm$9   & 3.94$\pm$0.89\\
\tableline
   \multicolumn{2}{l}{Normal ($M_{V}$ $<$ -6)}                 \\
     All Halo & 39 & 37$\pm$44  & 139$\pm$16 & 0.26$\pm$0.32\\
     YH       & 17 & -69$\pm$81 & 160$\pm$27 & -0.44$\pm$0.51\\
     OH       & 22 & 105$\pm$42 & 103$\pm$15 & 1.02$\pm$0.43\\
     D/B      & 11 & 195$\pm$27 & 52$\pm$11  & 3.76$\pm$0.95\\
\hline
\tablenotetext{*}{For $R_{gc}$ $<$ 40 kpc and excluding GCs with (cos$\psi$) $>$ 0.2}
\end{tabular}
\end{center}
\end{table}

The result of the kinematic analysis based on the constant-rotational-velocity solutions (\citealp{Zinn93}; \citealp{FW80}) 
and the updated database of Harris (\citealp{Harris96}) is presented in Table 1. When all the GCs are considered, 
we are basically confirming the conclusion of the previous work (\citealp{Zinn93}). YH group is dominated by random motion 
with no sign of significant rotation ($V_{rot}$), while OH group shows some prograde rotation and a smaller 
line-of-sight velocity dispersion ($\sigma_{los}$). D/B group is mostly supported by rotation with a relatively 
small $\sigma_{los}$. We find, however, EHB GCs, both belonging to YH and OH groups, are dominated by random motion 
and show no signs of rotation. Consequently, when they are excluded from the sample, normal GCs in OH group show 
increased rotation (from 1.5 to 1.8$\sigma$ from zero $V_{rot}$) and higher value of $V_{rot}$/$\sigma_{los}$. The same trend is also 
observed in the normal GCs in D/B group, but with much larger uncertainty. When only comparably bright ($M_{V}$ $<$ -6) 
GCs are considered, the differences become significantly larger (2.5$\sigma$ from zero $V_{rot}$). 
The above analysis, based only on the radial velocity data, provides good reason to suspect that EHB GCs are 
kinematically decoupled from other normal GCs, especially in OH group. Below, we investigate this in more detail 
using the measurements of full spatial motions and orbital parameters now available for 49 GCs in our sample (\citealp{Dinescu03}).

\begin{figure}[!b2]
\epsscale{0.95}
\plotone{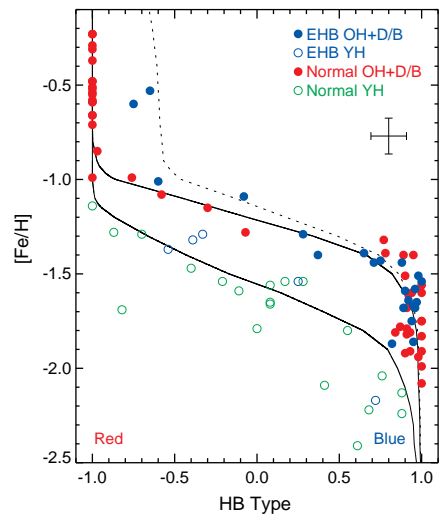}
\caption{The subdivision of GCs in the HB morphology versus metallicity diagram. The filled circles are GCs 
either in the ``old halo (OH)'' or metal-rich ([Fe/H] $>$ -0.8) ``disk/bulge (D/B)'' groups, while open circles are those in 
the ``younger halo (YH)'' group. The EHB GCs belong in all three subgroups, but most of them are in OH group. The 
updated database (Lee et al. 1994) consisting 95 GCs in $R_{gc}$ $<$ 40 kpc zone are compared with model HB isochrones (Rey et al. 2001)
in solid lines, the upper being older by 1.1 Gyr. Short dashed line has the same age as the upper solid line, 
but is for EHB GCs with 15 $\%$ of helium enhanced (Y = 0.33) subpopulation (\citealp{Lee05}).}
\end{figure}

\begin{figure*}
\epsscale{1.05}
\plotone{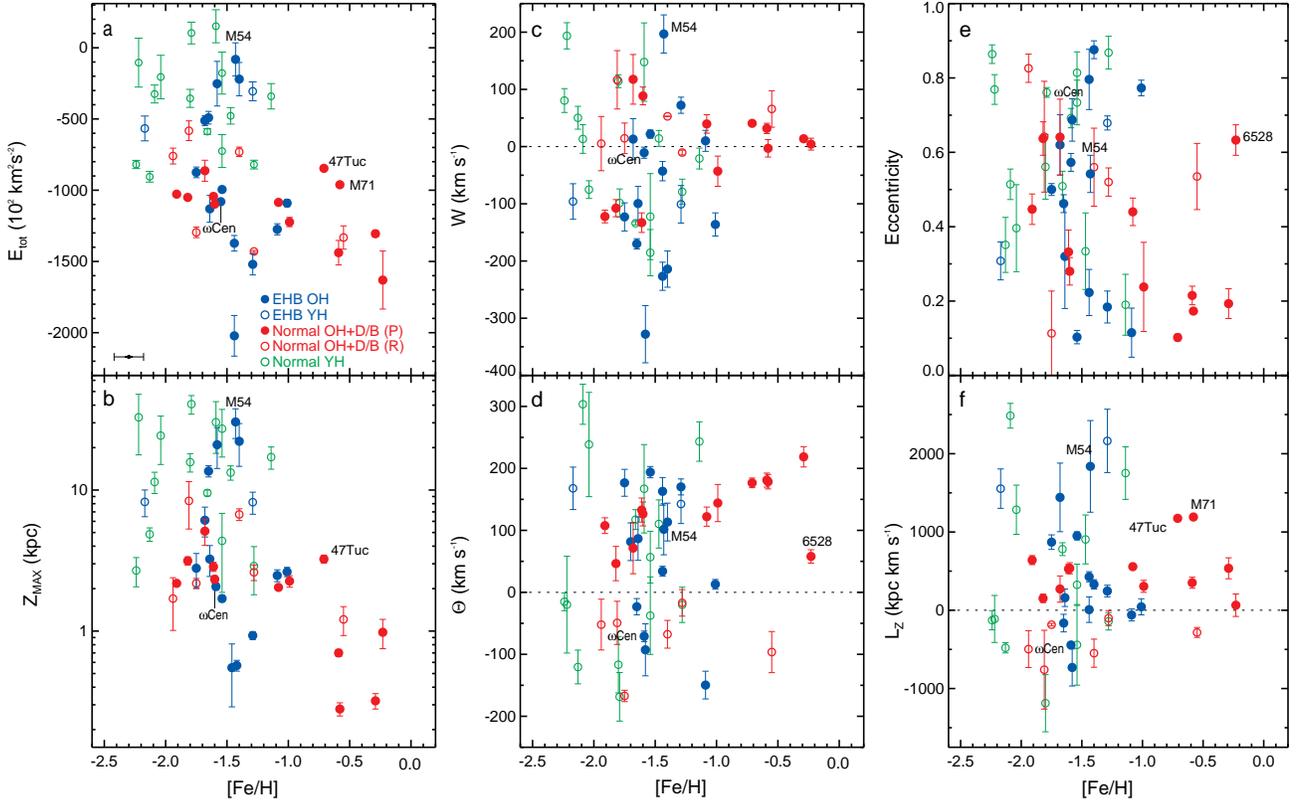}
\caption{The relationship between kinematics derived from full spatial motions and metallicity. From (a) to (f), 
total orbital energy ($E_{tot}$), maximum distance perpendicular to the Galactic plane ($Z_{max}$), velocity component perpendicular 
to the plane ($W$), rotational velocity ($\Theta$), orbital eccentricity, and the angular momentum component associated with $\Theta$ ($L_{Z}$) 
are plotted as a function of metallicity (Zinn 1993), respectively. Red filled circles are normal GCs with prograde rotation in 
OH and D/B groups. All of them are in the inner halo. Red open circles are normal GCs with retrograde rotation in OH and D/B groups. 
Only 3 of them are in the inner halo.}
\end{figure*}

In Figure 3, we have plotted kinematic parameters obtained from full spatial motions as a function of metallicity, which shows more 
directly the systematic differences between EHB and normal GCs. In all panels of Figure 3, EHB GCs have diversity in kinematics, 
and show no correlations with metallicity (correlation coefficient, \textit{r}, of -0.02 to -0.31 with high $p$-values of 0.25 to 0.94). 
Normal YH GCs, mostly in the outer halo, show kinematically hot signatures (high $E_{tot}$, $Z_{max}$, $\&$ eccentricity, 
and large velocity dispersion) (\citealp{MG04}). To our surprise, however, when EHB GCs are excluded, 
most normal GCs with prograde rotation in OH and D/B groups (red filled circles) show clear signs of dissipational collapse. 
$E_{tot}$, $Z_{max}$, $W$ velocity, and perhaps orbital eccentricity are all decreasing with increasing metallicity, 
among which the `chevron' shape of $W$ velocity distribution is most impressive. 
Rotational velocity, however, is increasing with metallicity, and $L_{Z}$ appears to be conserved. 
The orbital properties of \object{NGC 6528} are known to be highly affected by the potential of the bar 
because of its proximity (\citealp{Dinescu03}). Thus, excluding this one deviant point in panels (d) and (e), 
we obtain strong correlations for $E_{tot}$, $Z_{max}$, $|W|$, $\Theta$, and eccentricity (\textit{r} = -0.58, -0.71, -0.95, 
0.89, and -0.77, respectively) with small $p$-values (0.05, 0.01, 3.1$\times$ $10^{-6}$, 0.0002, and 0.005, respectively). 
In other words, the correlations are highly significant at the level of 95$\%$, 99$\%$, 99.999$\%$, 99.98$\%$, and 99.5$\%$, respectively. 
As expected, however, correlation is low (\textit{r} = 0.17) for $L_{Z}$ with a high $p$-value of 0.59.

All of these trends observed for normal GCs with prograde rotation in OH and D/B groups are fully consistent with the model first 
envisioned by Eggen, Lynden-Bell, $\&$ Sandage (1962), where metal enrichment went on as dissipational collapse continued. 
Although these results are based on relatively modest numbers of GCs with full spatial motion information, 
their coherent behaviours in all panels of Figure 3, together with statistically significant correlations, 
confirm that we are detecting real signatures. Also, these results are consistent with the kinematics solution 
obtained from radial velocity alone (Table 1), which is based on a larger sample of GCs. We argue, therefore, (1) 
EHB GCs in our sample are indeed kinematically decoupled from most of the normal GCs in OH group, and (2) when EHB GCs 
are excluded, we are detecting clearer signatures of dissipational collapse in the inner halo, which apparently led to 
the formation of the Galactic disk (\citealp{Zinn93}; \citealp{MG04}). The kinematics of EHB GCs, which are not following the dissipational collapse,
are more consistent with what one would expect among the relicts of primeval star-forming subsystems that first formed the 
nucleus (EHB GCs with low $E_{tot}$ and $Z_{max}$) and halo (EHB GCs with high $E_{tot}$ and $Z_{max}$) of the Galaxy through 
both dissipational and dissipationless mergers, as has been predicted by recent $\Lambda$CDM simulations 
for ``high-$\sigma$ peaks'' (e.g., Diemand, Madau, $\&$ Moore 2005; \citealp{Moore06}). As described above, 
a significant fraction of the helium enriched subpopulation also favours building block origin of 
EHB GCs. Normal YH GCs in the outer halo 
share their kinematic properties with the outlying EHB GCs, which is consistent with the view (\citealp{SZ78}) that 
they were originally formed in the outskirts of isolated building blocks and later accreted to the outer halo of the Galaxy 
when their parent dwarf galaxies, like the Sagittarius, were merging with the Milky Way. The GCs with EHB 
also tend to show more extended Na-O and Mg-Al anticorrelations (\citealp{Gratton07}). Therefore, 
the suggested connection between some of these GCs with strong chemical inhomogenity and orbital parameters (\citealp{Carretta06}) 
might be due to the diversity of kinematics among EHB GCs.

According to the present picture, most of the normal GCs with retrograde rotation in OH and D/B groups could have also 
originated from the subsystems  with retrograde rotation. Interestingly, their relatively confined distributions both in 
the angular momentum phase space (\citealp{Helmi99}) and velocity space are not inconsistent 
with the possibility that some or most of them were former members of parent dwarf galaxies hosting two EHB GCs, \object{$\omega$Cen} and/or 
\object{NGC 6723} (Lee et al., in preparation). Their distribution in velocity space is also well consistent with the model prediction of the tidal debris from $\omega$Cen's 
parent dwarf system (Mizutani, Chiba, $\&$ Sakamoto 2003), which was presumably formed in the outer halo and accreted to the inner halo. 
Note that a similar minor merging of subsystem with the thin disk (Quinn, Hernquist, $\&$ Fullagar 1993) could have also 
changed some of the original kinematic properties of two disk GCs in Figure 3 (\object{47Tuc} and \object{M71}).

\section{DISCUSSION}

The clear differences in kinematics and mass between GCs with and without EHB are strong evidence that they have different origins. 
Our results suggest present-day Galactic GCs are most likely an ensemble of heterogeneous objects originated from 
three distinct phases of the Milky Way formation: (1) remaining cores or central star clusters of building blocks that first assembled to form the 
nucleus and halo of the proto-Galaxy (\citealp{BC02}; \citealp{Santos03}; 
\citealp{Bekki05}; \citealp{KG05}; \citealp{Moore06}), (2) genuine GCs formed in the dissipational collapse of a transient gas-rich inner 
halo system that eventually formed the Galactic disk (\citealp{Eggen62}), and (3) genuine GCs formed in the outskirts of
outlying building blocks that later accreted to the outer halo of the Milky Way (\citealp{SZ78}). In this picture, 
relicts of first building blocks that formed the flattened nucleus (\citealp{KG05}; \citealp{Moore06}) are now 
observed as relatively metal-poor EHB GCs (e.g., NGC \object{6266}, \object{6522}, and \object{6626}) having low $E_{tot}$ and $Z_{max}$ near the centre. 
Formation of the slowly-rotating gas-rich inner halo system that later collapsed in phase (2) is still most unclear, 
but it is attractive to speculate that leftover gas from ``rare peaks'' (building blocks hosting EHB GCs) in the inner 
halo and gas from continuously falling ``less rare peaks'' (\citealp{Moore06}) led to the formation of this structure, 
perhaps with the aids of some heating feedbacks (e.g., \citealp{Schawinski06}) soon followed by cooling. 
Several lines of further study will certainly help to shed more light into the picture briefly sketched here. 
For example, search for the tidal streams that might be associated with EHB GCs, dark matter search in the outlying 
EHB GCs where preferential disruption of dark matter halo (\citealp{Saitoh06}) might be less severe, kinematics 
analyses of extragalactic GC systems along with the ultraviolet survey for EHB GC candidates, together 
with more detailed high resolution $\Lambda$CDM simulations.\\

We thank R. Zinn, R. Larson, and P. Demarque for helpful discussions, C. Chung for his assistance in HB isochrone construction, 
and H.-Y. Lee for her assistance in CMD compilation. Support for this work was provided by the Creative Research Initiatives 
Program of the Korean Ministry of Science $\&$ Technology and KOSEF, for which we are grateful.


\begin{thebibliography} {99}

\bibitem[Bedin et al. 2004]{Bedin04}  
Bedin, L. R., Piotto, G., Anderson, J., Cassisi, S., King, I. R., Momany, Y., \& 
      Carraro, G. 2004, \apj, 605, L125

\bibitem[Bekki 2005] {Bekki05}
Bekki, K. 2005, \apj, 626, L93

\bibitem[Bekki \& Norris 2006] {BN06}
Bekki, K. \& 
      Norris, J. E. 2006, \apj, 637, L109

\bibitem[Bromm \& Clarke 2002] {BC02}
Bromm, V. \&
      Clarke, C. J. 2002, \apj, 566, L1

\bibitem[Carretta 2006] {Carretta06}
Carretta, E. 2006, \aj, 131, 1766

\bibitem[D'Antona et al. 2005]{Dan05}
D'Antona, F., Bellazzini, M., Caloi, V., Pecci, F. F., Galleti, S., \& 
      Rood, R. T. 2005, \apj, 631, 868

\bibitem[Diemand et al. 2005]{Diemand05}
Diemand, J., Madau, P., \&
      Moore, B. 2005, \mnras, 364, 367

\bibitem[Dinescu et al. 2003]{Dinescu03}
Dinescu, D. I., Girard, T. M., van Altena, W. F., \& 
      Lopez, C. E. 2003, \aj, 125, 1373

\bibitem[Eggen et al. 1962]{Eggen62}
Eggen, O. J., Lynden-Bell, D., \& 
      Sandage, A. R. 1962, \apj, 136, 748

\bibitem[Freeman 1993]{Freeman93}
Freeman, K. C. 1993, in ASP Conf. Ser. 48, The Globular Clusters-Galaxy Connection, ed. G. H. Smith \& J. P. Brodie (San Francisco: ASP), 608

\bibitem[Frenk \& White 1980]{FW80}
Frenk, C. S. \& 
      White, S. D. M. 1980, \mnras, 193, 295

\bibitem[Gratton 2007] {Gratton07}
Gratton, R. 2007, in ASP Conf. Ser., From the Stars to Galaxies, ed. A. Vallenari \& R. Tantalo, (Venice: ASP), in press

\bibitem[Harris 1996]{Harris96}
Harris, W. E. 1996, \aj, 112, 1487

\bibitem[Helmi et al. 1999]{Helmi99}
Helmi, A., White, S. D. M., de Zeeuw, P. T., \& 
      Zhao, H. 1999, \nat, 402, 53

\bibitem[Kravtsov \& Gnedin 2005] {KG05}
Kravtsov, A. V. \& 
      Gnedin, O. Y. 2005, \apj, 623, 650

\bibitem[Layden \& Sarajedini 2000]{LS00}
Layden, A. C. \& 
      Sarajedini, A. 2000, \aj, 119, 1760

\bibitem[Lee, Demarque, \& Zinn 1994]  {Lee94}
Lee, Y.-W., Demarque, P., \&
      Zinn, R. 1994, \apj, 423, 248

\bibitem[Lee et al. 1999]{Lee99}
Lee, Y.-W., Joo, J.-M., Sohn, Y.-J., Rey, S.-C., Lee, H.-C., \& 
      Walker, A. R. 1999, \nat, 402, 55

\bibitem[Lee et al. 2005]{Lee05}
Lee, Y.-W. et al. 2005, \apj, 621, L57

\bibitem[Mackey \& Gilmore 2004]{MG04}
Mackey, A. D. \& 
      Gilmore, G. F. 2004, \mnras, 355, 504

\bibitem[Mizutani, Chiba, \& Sakamoto 2003]{MCS03}
Mizutani, A., Chiba, M., \& 
      Sakamoto, T. 2003, \apj, 589, L89

\bibitem[Moore et al. 2006]{Moore06}
Moore, B., Diemand, J., Madau, P., Zemp, M., 
      Stadel, J. 2006, \mnras, 368, 563

\bibitem[Norris 2004]{Norris04}
Norris, J. E. 2004, \apj, 612, L25

\bibitem[Piotto et al. 2002]{Piotto02}
Piotto, G. et al. 2002, \aap, 391, 945

\bibitem[Piotto et al. 2005]{Piotto05}
Piotto, G. et al. 2005, \apj, 621, 777

\bibitem[Piotto et al. 2007]{Piotto07}
Piotto, G. et al. 2007, preprint (astro-ph/0703767)

\bibitem[Quinn, Hernquist, \& Fullagar 1993]{QHF93}
Quinn, P., Hernquist, L, \& 
      Fullagar, D. 1993, \apj, 430, 74

\bibitem[Recio-Blanco et al. 2006]{RB06}
Recio-Blanco, A., Aparico, A., Piotto, G., de Angeli, F., \& 
      Djorgovski, S. J. 2006, \aap, 452, 875

\bibitem[Rey et al. 2001]{Rey01}
Rey, S.-C., Yoon, S. J., Lee, Y.-W., Chaboyer, B., \& 
      Sarajedini, A. 2001, \aj, 122, 3219

\bibitem[Rosenberg, Recio-Blanco, \& Garcia-Marin 2004]{Rosen04}
Rosenberg, A., Recio-Blanco, A., \&
      Garcia-Marin, M. 2004, \apj, 603, 135

\bibitem[Saitoh et al. 2006]{Saitoh06}
Saitoh, T. R., Koda, J., Okamoto, T., Wada, K., \& 
      Habe, A. 2006, \apj, 640, 22

\bibitem[Salaris \& Weiss 2002]{SW02}
Salaris, M. \& 
      Weiss, A. 2002, \aap, 388, 492

\bibitem[Santos 2003] {Santos03}
Santos, M. R. 2003, in ESO Astrophysics Symposia, Extragalactic Globular Cluster Systems, ed. M. Kissler-Patig (Garching: Springer-Verlag), 348

\bibitem[Schawinski et al. 2006]{Schawinski06}
Schawinski, K. et al., 2006, \nat, 442, 888

\bibitem[Searle \& Zinn 1978]{SZ78}
Searle, L \&
      Zinn, R. 1978, \apj, 225, 357

\bibitem[Zinn 1993]{Zinn93}
Zinn, R. 1993, in ASP Conf. Ser. 48, The Globular Clusters-Galaxy Connection, ed. G. H. Smith \& J. P. Brodie (San Francisco: ASP), 38

\end{thebibliography}
\end{document}